# Graphene-Passivated Nickel as an Oxidation-Resistant Electrode for Spintronics

Bruno Dlubak[1], Marie-Blandine Martin[2], Robert S. Weatherup[1], Heejun Yang[2], Cyrile Deranlot[2], Raoul Blume[3], Robert Schloegl[4], Albert Fert[2], Abdelmadjid Anane[2], Stephan Hofmann[1], Pierre Seneor[2], and John Robertson[1]

1. Department of Engineering, University of Cambridge, Cambridge CB2 1PZ, United Kingdom

2. Unité Mixte de Physique CNRS/Thales, 91767 Palaiseau, France and University of Paris-Sud, 91405 Orsay, France

3. Helmholtz-Zentrum Berlin fur Materialien und Energie, 12489 Berlin, Germany

4. Department of Inorganic Chemistry, Fritz Haber Institute of the Max Planck Society, Faradayweg 4-6, 14195 Berlin, Germany

**We report on graphene-passivated ferromagnetic electrodes (GPFE) for spin devices. GPFE are shown to act as spin-polarized oxidation-resistant electrodes. The direct coating of nickel with few layer graphene through a readily scalable chemical vapour deposition (CVD) process allows the preservation of an unoxidized nickel surface upon air exposure. Fabrication and measurement of complete reference tunneling spin valve structures demonstrates that the GPFE is maintained as a spin polarizer and also that the presence of the graphene coating leads to a specific sign reversal of the magneto-resistance. Hence, this work highlights a novel oxidation-resistant spin source which further unlocks low cost wet chemistry processes for spintronics devices.**

Information storage is today mainly based on magnetism, with hundreds of millions of hard drives sold every year,[1,2] and further growth is expected driven by the proliferation of enormous data centers for online "cloud" computing. Spintronics is at the heart of this non-volatile data-storage revolution, with ferromagnetic memory elements acting as basic building blocks: spin sources or analyzers. The efficiency of spin polarized electrodes, based on ferromagnetic metals like nickel and cobalt, thereby heavily relies on the quality of the interfaces at play. A major challenge for device processing and integration is to prevent detrimental corrosion and oxidation of the ferromagnetic metals in use. To date, this severely limits the use of ambient and low cost wet processing steps and the integration of novel materials like organic molecules and chemically derived nanostructures in spintronics.

Here we present a simple, scalable process to fabricate oxidation-resistant spin polarized electrodes that we anticipate can open up a range of new integrative pathways for spin devices. Our process uses the ferromagnetic metal, here Ni, as catalyst for the low temperature chemical vapour deposition (CVD) of graphene layers.[3-5] The CVD process ensures the reduction of the Ni surface and the resulting graphene coating acts as an oxidation passivation. Importantly this graphene passivated ferromagnetic electrode (GPFE) preserves a spin polarization for electrons flowing perpendicularly through it. We characterize the GPFE by *in-situ* X-ray photoelectron spectroscopy (XPS). Furthermore, we identify a particular filtering of majority spins by the GPFE, through magneto-transport measurements of a complete tunneling spin valve structure which reveal a negative magneto-resistance MR = -10.8%. Graphene has already been identified as a promising material for current-perpendicular-to-plane (CPP) spintronics devices, but previous studies relied on graphene exfoliation[6,7] or transfer[8] and hence could not utilize the gas impermeability of the graphene layers.[9,10] We establish here a process technology for the direct integration of graphene in to spintronics devices, that fully makes use of this key advantage of graphene.

**RESULTS AND DISCUSSION**

Figure 1 shows the principle of our process and the device lay-out. Lithographically defined Ni stripes on $SiO_2$/Si support are exposed to a hydrocarbon in a one-step CVD process at 600°C. This results in a selective, conformal coating of the Ni with few layer graphene (FLG ~ 2-5 layers). We previously reported[5,11] on all details of low temperature, Ni catalyzed graphene CVD and the related growth mechanisms. Importantly, the graphene layer is thus directly grown on the structure without the need for the usual transfer steps involved in the fabrication of CVD graphene devices,[8] and the process is readily scalable.

As highlighted earlier, the passivation of the nickel surface is crucial as, once exposed to ambient air, a bare nickel surface is immediately oxidized. Hence, depositing graphene on an air exposed nickel electrode by exfoliation of graphite or by transfer of CVD grown graphene may lead to an undesired and ill-defined Ni/NiOx/graphene electrode, where the NiOx layer acts on its own as a tunnel barrier with a poorly defined effect on spin properties.[12,13] We therefore use *in-situ* X-ray photoelectron spectroscopy (XPS) measurements to confirm that our approach of direct graphene CVD on nickel at 600 °C results in an oxide-free GPFE. $Ni2p_{3/2}$ core level spectra were acquired at several stages of the CVD process, and following an extended exposure to air after the CVD process (Fig. 2). Prior to annealing, the measured $Ni2p_{3/2}$ spectrum of the surface of the as-deposited Ni layer is characteristic of oxidized Ni (Fig. 2, top spectrum). After heating to 300 °C in a $H_2$ atmosphere, the XPS oxide peaks are completely removed and the peaks at 852.6 eV ($Ni_M$) and 853.0 eV ($Ni_{Dis}$) become dominant (Fig. 2, middle spectrum), as expected for oxide-free metallic Ni.[5,11] The observation of this characteristic metallic Ni spectrum confirms that the Ni surface is completely reduced during the annealing step of the CVD process. The temperature is further raised to 600 °C (without significant change in the XPS spectra) and the FLG is then grown by CVD (see methods). After cooling, the graphene-passivated Ni sample is transferred in ambient air in order to proceed to following lithographic steps. To emphasize the protection effect of the graphene layer, a reference sample was produced and left exposed to ambient atmosphere for 7 days. The subsequent XPS measurements of the the Ni2p$_{3/2}$ core level (Fig. 2, bottom spectrum) shows that the Ni$_M$ and Ni$_{Dis}$ components are still dominant and no further peaks have emerged (*i.e.* the spectra features are unchanged from the *in-situ* spectra after reduction) indicating that Ni remains reduced even after extended exposure to air. We note the intensity of the lower spectrum is much less than for the other spectra, as a result of the Ni being covered with FLG. While pristine monolayer graphene acts as an impermeable membrane even to He,[9] the presence of defects in the sp$^2$ structure may be thought to provide paths for gas diffusion. However, the CVD process in use here (see methods) leads to a self-terminating FLG film which acts as an effective oxygen diffusion barrier under ambient conditions as shown by XPS, and thus protects the GPFE from oxidation.

To characterize the electronic transport properties of the GPFE, a graphene-nickel stripe is contacted with a reference Al$_2$O$_3$/Co probe structure (Fig. 1b). To achieve this, first a 1 μm × 1 μm square is opened in a resist above the graphene coated Ni stripe. The Al$_2$O$_3$(1 nm)/Co(15 nm)/Au top contact structure is then produced by sputtering. In particular the Al$_2$O$_3$ layer is deposited in two steps: a 0.6 nm Al layer is sputtered, and is then further oxidized in 50 Torr O$_2$ atmosphere leading to a homogenous 1 nm Al$_2$O$_3$ film on graphene.[14] The resulting structure is composed of the GPFE with the spin polarized tunneling current probe on top over a 1 μm$^2$ area (Fig. 1b).

Figures 3 and 4 present the characterization of the GPFE/probe electrical properties at 1.4K. The measured resistance × area product of the structure is in the *MΩ.μm$^2$* range. This is in agreement with previous characterization of the sputtered 1 nm Al$_2$O$_3$ tunnel barrier on graphene.[14] *dI/dV* spectroscopy characterization of the junction is carried with an AC+DC lock-in based measurement setup. The *dI/dV* tunneling spectrum presented in Fig. 3a reveals a ~120 meV wide gap-like feature at the Fermi level ($E_F$). This gap is a characteristic signature of electrons tunneling into graphene, as revealed previously in STM studies for graphene on SiC,[15] SiO$_2$,[16] BN,[17] and Pt,[18] and described by *ab-initio* calculation of the tunnelling density of states.[19] Indeed, at low bias (*< 60 mV*), only elastic tunneling paths are enabled near the graphene's K points at $E_F$. Due to the particular band structure of graphene this leads to a quenching of the injected current (Fig. 3b) due to a *k* vector mismatch: the current in graphene is carried by electron's having non-zero in-plane momentum, $k_=$, while the distribution of tunneling probabilities through our *1 nm* Al$_2$O$_3$ tunneling layer is maximum for $k_= = 0$ and presents an exponential decay with increasing $k_=$ as shown in STM-tip/graphene measurements.[15-19] This is emphasized in our measurements by the fact that when energies ascribed to out-of-plane acoustic graphene phonon mode at *≈ 60 meV* are reached, [15-19] additional inelastic tunneling paths are activated and the current rises. The suppression of the tunneling for small biases and the identification of the phonon-mediated inelastic tunneling channels in the *dI/dV* spectroscopy further show, in addition to XPS measurements, that the transport occurs as expected in a well-defined Ni/graphene/Al$_2$O$_3$/Co structure.

Figure 4 presents magneto-dependent measurements through the junction. While both spin polarizations of Co/Al$_2$O$_3$ and Ni/Al$_2$O$_3$ interfaces are known to be positive [20,21] and hence lead to positive magneto-resistance signals in Ni/Al$_2$O$_3$/Co structures [22,23] (see Fig. 5a), a *negative* magneto-resistance is observed in our system due to the simple insertion of the graphene layer between the Ni electrode and the Al$_2$O$_3$ tunnel barrier. Following De Teresa *et* *al.*[24] and taking the definition $MR = \frac{R_{AP}-R_P}{R_{AP}} = \frac{2P_{GPFE}\,P_{SP}}{1+P_{GPFE}\,P_{SP}}$, we find $MR = -10.8\%$ in our system (Fig. 4a) from which we derive for the nickel-GPFE a large negative spin polarization, *$P_{GPFE}$ = -16%* (see Ref. [25]), by assuming *$P_{SP}$ = +32%* for the Co/Al$_2$O$_3$ spin probe. This estimation of $P_{GPFE}$ is thus a lower bound of the spin polarization amplitude as we take the maximum value of $P_{SP}$ extracted from previous devices.[14,26,27] While spin polarizations reported for electron tunneling from Ni/Al$_2$O$_3$ electrode are positive in complete TMR structures [22,23] as well as in ferromagnet/insulator/superconductor tunneling structures,[20,21] we observe a drastic evolution of the spin polarization which notably leads to a sign reversal of the Ni spin polarization at the GPFE. We consistently observed negative magnetoresitance signals (-5% to -10%) among different runs and positive magnetoresistance was never observed.

This shows that as conceptually expected for molecules [28] and as predicted from *ab-initio* calculations for the GPFE,[29] the sole presence of the graphene passivation layer is able to induce a spin filtering effect and the reversal of the spin polarization. This result can be simply understood in terms of filtering of Ni majority spins by graphene. Indeed, when comparing the Fermi surface of graphene to the one of nickel, it appears that at graphene's K points nickel presents only minority spin electrons:[29] minority spins thus have a continuous transport channel through the GPFE, while majority spins have no direct conduction path and are filtered out (Fig. 5b).

**CONCLUSION**

In conclusion, we show that spin polarized ferromagnetic electrodes can be successfully passivated against oxidation by the growth of graphene by CVD. This is especially interesting when targeting organic-based spintronics devices where oxidation and other chemical reactions occurring at interfaces could quench spin signals. The potential of GPFEs for organic-based electronics has also been previously highlighted by the observed radical enhancement of the wetting ability of metal/graphene systems [30] which is envisioned to also assist their coating with organic molecules like pentacene or phthalocyanines.[31,32] Furthermore, experimental evidence presented here confirms that a particular spin-filtering effect takes place at the GPFE, with the graphene layer shown to drastically modify the spin polarization properties of a ferromagnetic metal, leading to spin polarization reversal. The growth of high-quality CVD graphene on Ni based catalysts at CMOS-compatible temperatures (< 450 °C) has been previously demonstrated[11] and thus the direct incorporation of GPFE spin filters in integrated spintronics devices can be envisioned. Finally, the presented results in this paper highlight a possible path to the all-spin logic device described by Behin-Aein *et al.*,[33] using graphene as a global platform for spin processing architectures, where it could both translate magnetically stored information from nickel dot registers into the corresponding electron's spin polarization, and further transport this spin information with a high efficiency.[14]

**METHODS**

The GPFE geometry (Figure 1a) was defined by ebeam lithography using Shipley's UVIII resist on a $SiO_2$(300nm)/Si substrate. A 10 μm wide stripe was opened in the resist, a 150 nm thick nickel layer was then deposited by evaporation and a standard lift-off step was carried. Following previous reports,[5,11] the nickel was then covered with graphene through a chemical vapour deposition growth step in a custom-built cold-wall reactor whose base pressure is $5\times10^{-7}$ mbar. The sample was heated up to 600°C at about 300 °C/min and annealed in a 1 mbar atmosphere of $H_2$ for 15 minutes. The $H_2$ was removed and then the sample was exposed to a $10^{-5}$ mbar atmosphere of $C_2H_2$ at 600°C for 15 minutes. Finally, the sample was cooled in vacuum at ~100 °C/min. This leads to complete coverage of nickel by a FLG film. No significant increase in average layer number is observed for longer exposure times. We attribute this self-terminating growth to the grown layers blocking the precursor supply to the Ni catalyst.[5]

The spin probe analyzer electrode geometry was defined in a second ebeam step. A 1 μm x 1 μm window was opened in UVIII resist on top of the GPFE. The stack was then deposited by sputtering, through a shadow mask which protected bonding pads. First, a 0.6 nm thick layer of aluminum was sputtered by DC plasma, followed by exposure to a 50 Torr $O_2$ atmosphere for 10 minutes. A 15 nm cobalt layer was then sputtered on top and capped by a 80 nm gold layer. The spin properties of this spin probe have previously been studied [14,26,27] with a maximum measured spin polarization of $P_{SP}=+32\%$.

*In situ* XPS measurements revealing the reduction and the passivation of the nickel layer during low-pressure CVD were performed at the BESSY II synchrotron at the ISISS end station of the FHI-MPG. An IR laser focused onto a SiC backplate was used for sample heating. Temperature readings were taken from a thermocouple clamped to the sample surface close to the measured region, and as such, this may lead to an uncertainty in the actual sample temperature of ~50 °C.

# REFERENCES

1. Wolf, S. A.; Awschalom, D. D.; Buhrman, R. A.; Daughton, J. M.; Molnár, S. von; Roukes, M. L.; Chtchelkanova, A. Y.; Treger, D. M. Spintronics: A Spin-Based Electronics Vision for the Future. *Science* **2001**, *294*, 1488-1495.

2. Chappert, C.; Fert, A.; Dau, F. N. Van The Emergence of Spin Electronics in Data Storage. *Nat. Mater.* **2007**, *6*, 813-823.

3. Yu, Q.; Lian, J.; Siriponglert, S.; Li, H.; Chen, Y. P.; Pei, S.-S. Graphene Segregated on Ni Surfaces and Transferred to Insulators. *Appl. Phys. Lett.* **2008**, *93*, 113103.

4. Reina, A.; Jia, X.; Ho, J.; Nezich, D.; Son, H.; Bulovic, V.; Dresselhaus, M. S.; Kong, J. Large Area, Few-Layer Graphene Films on Arbitrary Substrates by Chemical Vapor Deposition. *Nano Lett.* **2009**, *9*, 30-35.

5. Weatherup, R. S.; Bayer, B. C.; Blume, R.; Baehtz, C.; Kidambi, P. R.; Fouquet, M.; Wirth, C. T.; Schlögl, R.; Hofmann, S. On the Mechanisms of Ni-Catalysed Graphene Chemical Vapour Deposition. *ChemPhysChem* **2012**, *13*, 2544–2549.

6. Mohiuddin, T. M. G.; Hill, E.; Elias, D.; Zhukov, A.; Novoselov, K.; Geim, A. Graphene in Multilayered CPP Spin Valves. *IEEE Trans. Magn.* **2008**, *44*, 2624-2627.

7. Banerjee, T.; Wiel, W. G. van der; Jansen, R. Spin Injection and Perpendicular Spin Transport in Graphite Nanostructures. *Phys. Rev. B* **2010**, *81*, 214409.

8. Cobas, E.; Friedman, A. L.; Van't Erve, O. M. J.; Robinson, J. T.; Jonker, B. T. Graphene as a Tunnel Barrier: Graphene-Based Magnetic Tunnel Junctions. *Nano Lett.* **2012**, *12*, 3000-3004.

9. Bunch, J. S.; Verbridge, S. S.; Alden, J. S.; Zande, A. M. van der; Parpia, J. M.; Craighead, H. G.; McEuen, P. L. Impermeable Atomic Membranes from Graphene Sheets. *Nano Lett.* **2008**, *8*, 2458-2462.

10. Chen, S.; Brown, L.; Levendorf, M.; Cai, W.; Ju, S.-Y.; Edgeworth, J.; Li, X.; Magnuson, C. W.; Velamakanni, A.; Piner, R. D. *et al.* Oxidation Resistance of Graphene-Coated Cu and Cu/Ni Alloy. *ACS Nano* **2011**, *5*, 1321-1327.

11. Weatherup, R. S.; Bayer, B. C.; Blume, R.; Ducati, C.; Baehtz, C.; Schloegl, R.; Hofmann, S. *In-Situ* Characterization of Alloy Catalysts for Low Temperature Graphene Growth. *Nano Lett.* **2011**, *11*, 4154-4160.

12. Maekawa, S.; Gafvert, U. Electron Tunneling Between Ferromagnetic Films. *IEEE Trans. Magn.* **1982**, *18*, 707-708.

13. Tsymbal, E. Y.; Sokolov, A.; Sabirianov, I. F.; Doudin, B. Resonant Inversion of Tunneling Magnetoresistance. *Phys. Rev. Lett.* **2003**, *98*, 186602.

14. Dlubak, B.; Martin, M.-B.; Deranlot, C.; Servet, B.; Xavier, S.; Mattana, R.; Sprinkle, M.; Berger, C.; Heer, W. A. De; Petroff, F. *et al.* Highly Efficient Spin Transport in Epitaxial Graphene on SiC. *Nat. Phys.* **2012**, *8*, 557-561.

15. Brar, V. W.; Zhang, Y.; Yayon, Y.; Ohta, T.; McChesney, J. L.; Bostwick, A.; Rotenberg, E.; Horn, K.; Crommie, M. F. Scanning Tunneling Spectroscopy of Inhomogeneous Electronic Structure in Monolayer and Bilayer Graphene on SiC. *Appl. Phys. Lett.* **2007**, *91*, 122102.

16. Zhang, Y.; Brar, V. W.; Wang, F.; Girit, C.; Yayon, Y.; Panlasigui, M.; Zettl, A.; Crommie, M. F. Giant Phonon-Induced Conductance in Scanning Tunnelling Spectroscopy of Gate-Tunable Graphene. *Nat. Phys.* **2008**, *4*, 627-630.

17. Decker, R.; Wang, Y.; Brar, V. W.; Regan, W.; Tsai, H.-Z.; Wu, Q.; Gannett, W.; Zettl, A.; Crommie, M. F. Local Electronic Properties of Graphene on a BN Substrate *via* Scanning Tunneling Microscopy. *Nano Lett.* **2011**, *11*, 2291-2295.

18. Levy, N.; Burke, S. A.; Meaker, K. L.; Panlasigui, M.; Zettl, A.; Guinea, F.; Neto, A. H. C.; Crommie, M. F. Strain-Induced Pseudo-Magnetic Fields Greater Than 300 Tesla in Graphene Nanobubbles. *Science* **2010**, *329*, 544-547.

19. Wehling, T.; Grigorenko, I.; Lichtenstein, A.; Balatsky, A. Phonon-Mediated Tunneling into Graphene. *Phys. Rev. Lett.* **2008**, *101*, 216803.

20. Monsma, D. J.; Parkin, S. S. P. Temporal Evolution of Spin-Polarization in Ferromagnetic Tunnel Junctions. *Appl. Phys. Lett.* **2000**, *77*, 883.

21. Miao, G.-X.; Münzenberg, M.; Moodera, J. S. Tunneling Path Toward Spintronics. *Rep. Prog. Phys.* **2011**, *74*, 036501.

22. Miyazaki, T.; Tezuka, N. Spin Polarized Tunneling in Ferromagnet / Insulator / Ferromagnet Junctions. *J. Magn. Magn. Mater.* **1995**, *151*, 403-410.

23. Suezawa, Y.; Takahashi, F.; Gondo, Y. Spin-Polarized Electron Tunneling in Ni/Al$_2$O$_3$/Co Junction and Large Magnetoresistance of Ni/Co Double Layers. *Jpn. J. Appl. Phys.* **1992**, *31*, L1415-L1416.

24. Teresa, J. De; Barthélémy, A.; Fert, A.; Contour, J.; Lyonnet, R.; Montaigne, F.; Seneor, P.; Vaurès, A. Inverse Tunnel Magnetoresistance in Co/SrTiO$_3$/La$_{0.7}$Sr$_{0.3}$MnO$_3$: New Ideas on Spin-Polarized Tunneling. *Phys. Rev. Lett.* **1999**, *82*, 4288-4291.

25. Jullière, M. Tunneling Between Ferromagnetic Films. *Phys. Lett.* **1975**, *54*, 225-226.

26. Bernand-Mantel, A.; Seneor, P.; Lidgi, N.; Munoz, M.; Cros, V.; Fusil, S.; Bouzehouane, K.; Deranlot, C.; Vaures, A.; Petroff, F. *et al.* Evidence for Spin Injection in a Single Metallic Nanoparticle: A Step Towards Nanospintronics. *Appl. Phys. Lett.* **2006**, *89*, 62502-62503.

27. Barraud, C.; Deranlot, C.; Seneor, P.; Mattana, R.; Dlubak, B.; Fusil, S.; Bouzehouane, K.; Deneuve, D.; Petroff, F.; Fert, A. Magnetoresistance in Magnetic Tunnel Junctions Grown on Flexible Organic Substrates. *Appl. Phys. Lett.* **2010**, *96*, 072502.

28. Barraud, C.; Seneor, P.; Mattana, R.; Fusil, S.; Bouzehouane, K.; Deranlot, C.; Graziosi, P.; Hueso, L.; Bergenti, I.; Dediu, V. *et al.* Unravelling the Role of the Interface for Spin Injection into Organic Semiconductors. *Nat. Phys.* **2010**, *6*, 615-620.

29. Karpan, V.; Giovannetti, G.; Khomyakov, P.; Talanana, M.; Starikov, A.; Zwierzycki, M.; Brink, J. van den; Brocks, G.; Kelly, P. Graphite and Graphene as Perfect Spin Filters. *Phys. Rev. Lett.* **2007**, *99*, 176602.

30. Dlubak, B.; Kidambi, P. R.; Weatherup, R. S.; Hofmann, S.; Robertson, J. Substrate-Assisted Nucleation of Ultra-Thin Dielectric Layers on Graphene by Atomic Layer Deposition. *Appl. Phys. Lett.* **2012**, *100*, 173113.

31. Roos, M.; Künzel, D.; Uhl, B.; Huang, H.-H.; Brandao Alves, O.; Hoster, H. E.; Gross, A.; Behm, R. J. Hierarchical Interactions and Their Influence Upon the Adsorption of Organic Molecules on a Graphene Film. *J. Am. Chem. Soc.* **2011**, *133*, 9208-9211.

32. Zhang, H.; Sun, J.; Low, T.; Zhang, L.; Pan, Y.; Liu, Q.; Mao, J.; Zhou, H.; Guo, H.; Du, S. *et al.* Assembly of Iron Phthalocyanine and Pentacene Molecules on a Graphene Monolayer Grown on Ru(0001). *Phys. Rev. B* **2011**, *84*, 245436.

33. Behin-Aein, B.; Datta, D.; Salahuddin, S.; Datta, S. Proposal for an All-Spin Logic Device with Built-In Memory. *Nat. Nano.* **2010**, *5*, 266-270.

**FIGURES CAPTIONS**

**Figure 1.** (a) Optical image of the device after graphene growth: a Ni stripe is coated with CVD graphene and a $Al_2O_3$/Co electrode is then deposited on lithographed 1 μm × 1 μm squares in UV resist. (b) Cross-sectional schematic of the junction

**Figure 2.** Fitted X-ray photoelectron spectra (XPS) of the nickel surface. Top XPS spectrum is characteristic of oxidized as-deposited nickel once exposed to air. Middle XPS spectrum reveals a reduced Ni surface after *in-situ* 300 °C treatment. Bottom XPS spectrum confirms the protection of the Ni surface by the CVD grown graphene layer: the preserved metallic nature of the nickel surface is probed through the graphene layer, even after a 7 days exposure to air.

**Figure 3.** (a) Tunnel spectroscopy of the Co/$Al_2O_3$/Graphene/Ni junction (2 sweeps). The observed ≈ 120 meV gap-like feature and the phonon-mediated activation of tunneling at larger biases are characteristic of electron tunneling perpendicularly to a graphene layer. Inset: non-linear I(V) trace. (b) Tunneling of $k_= = 0$ and $k_= = K$ electrons are both impeded at low biases.

**Figure 4.** Spin transport through the Ni/graphene/$Al_2O_3$/Co junction at 1.4K (a) A negative magnetoresistance MR = -10.8% is measured (applied DC bias 100 mV). (b) Magnetoresistance measured at -100 mV.

**Figure 5.** (a) A positive spin signal is expected in Ni/$Al_2O_3$/Co structures as reported in [22,23]. (b) Here, the measured negative spin signal is understood in terms of filtering of majority spins by the GPFE.

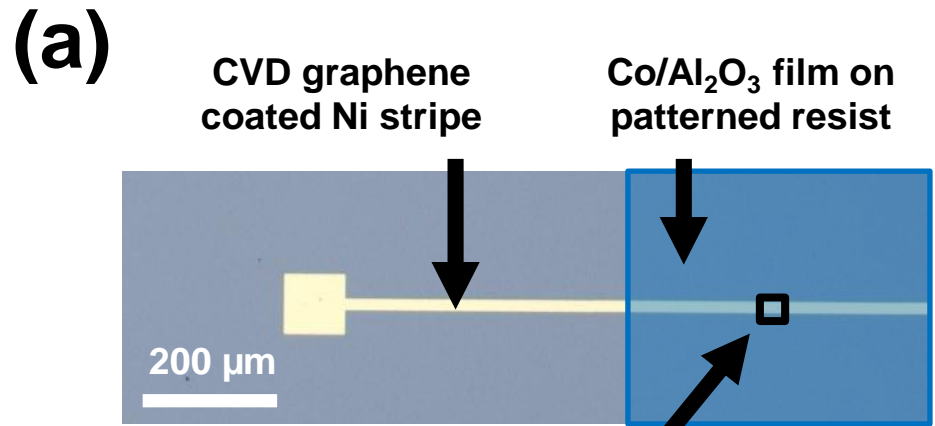
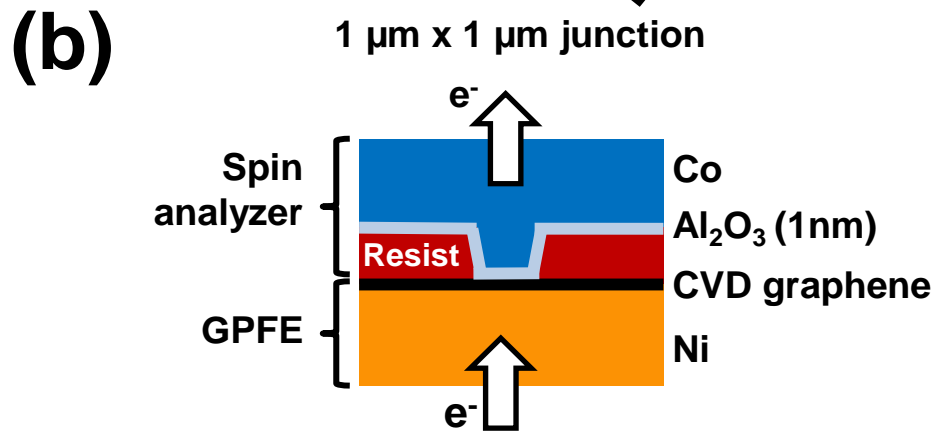

**Figure 1**

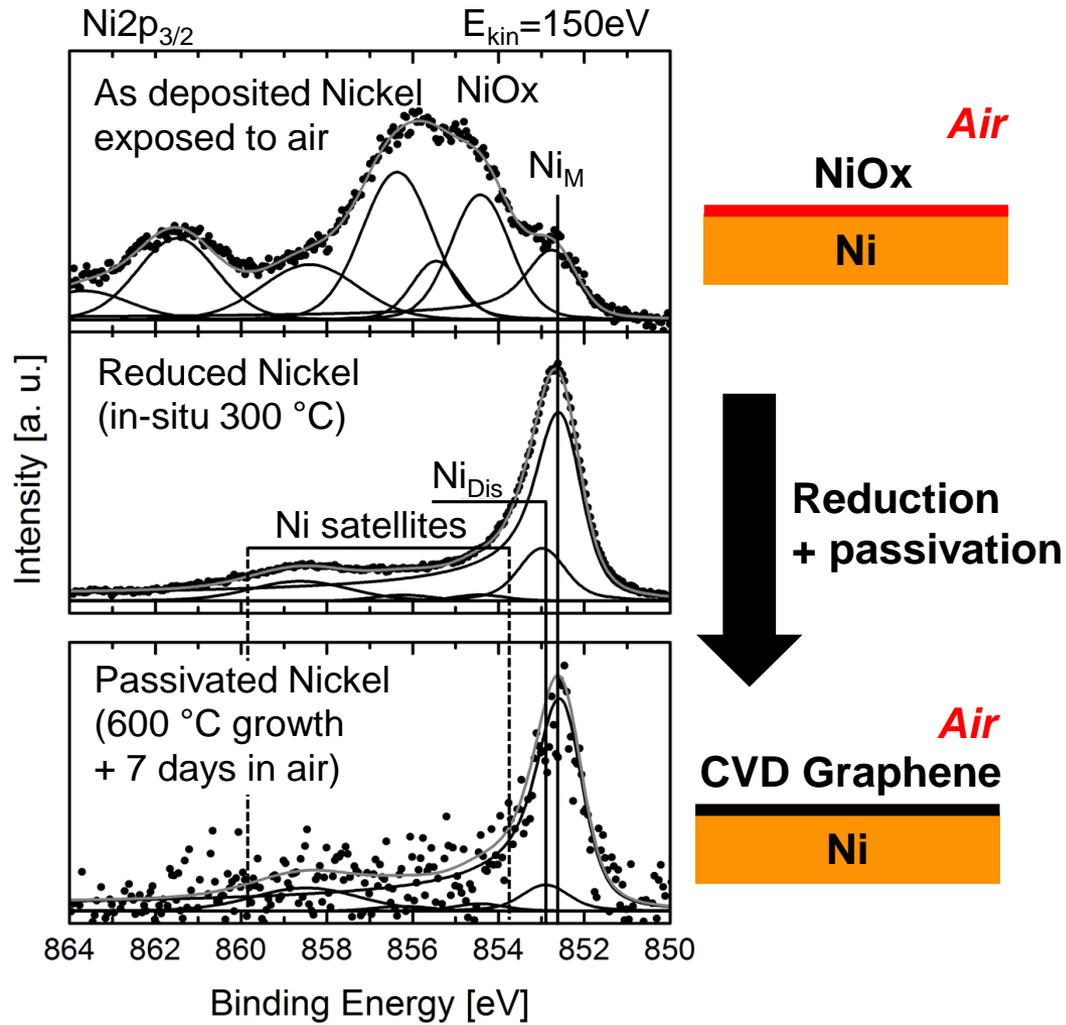

**Figure 2**

**(a)**

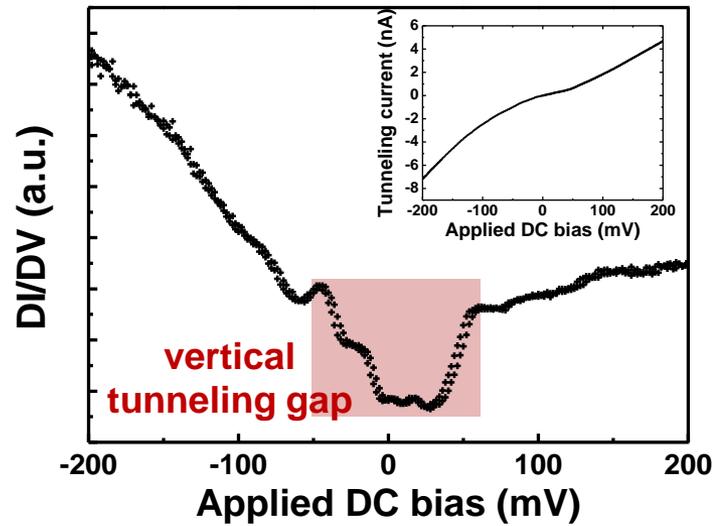

**(b)**

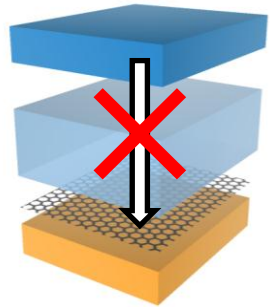
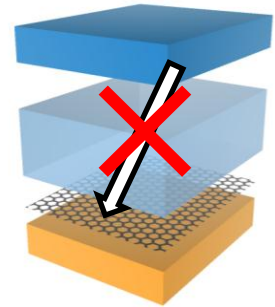

**Figure 3**

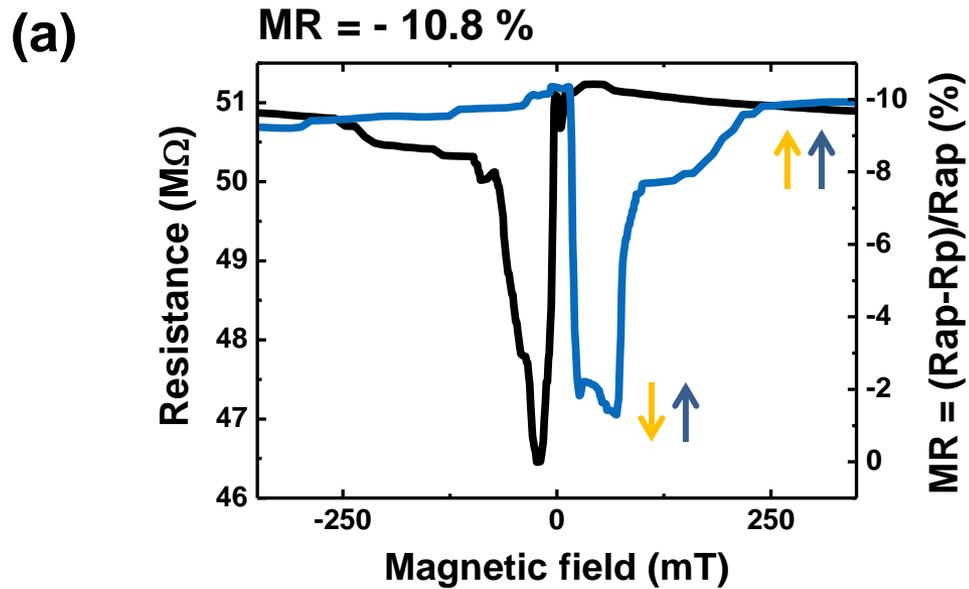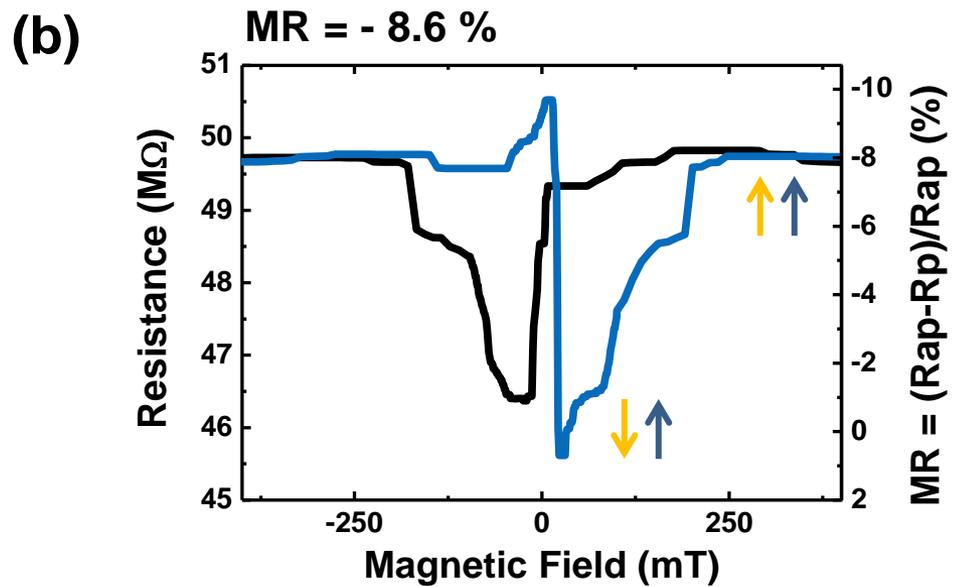

Figure 4

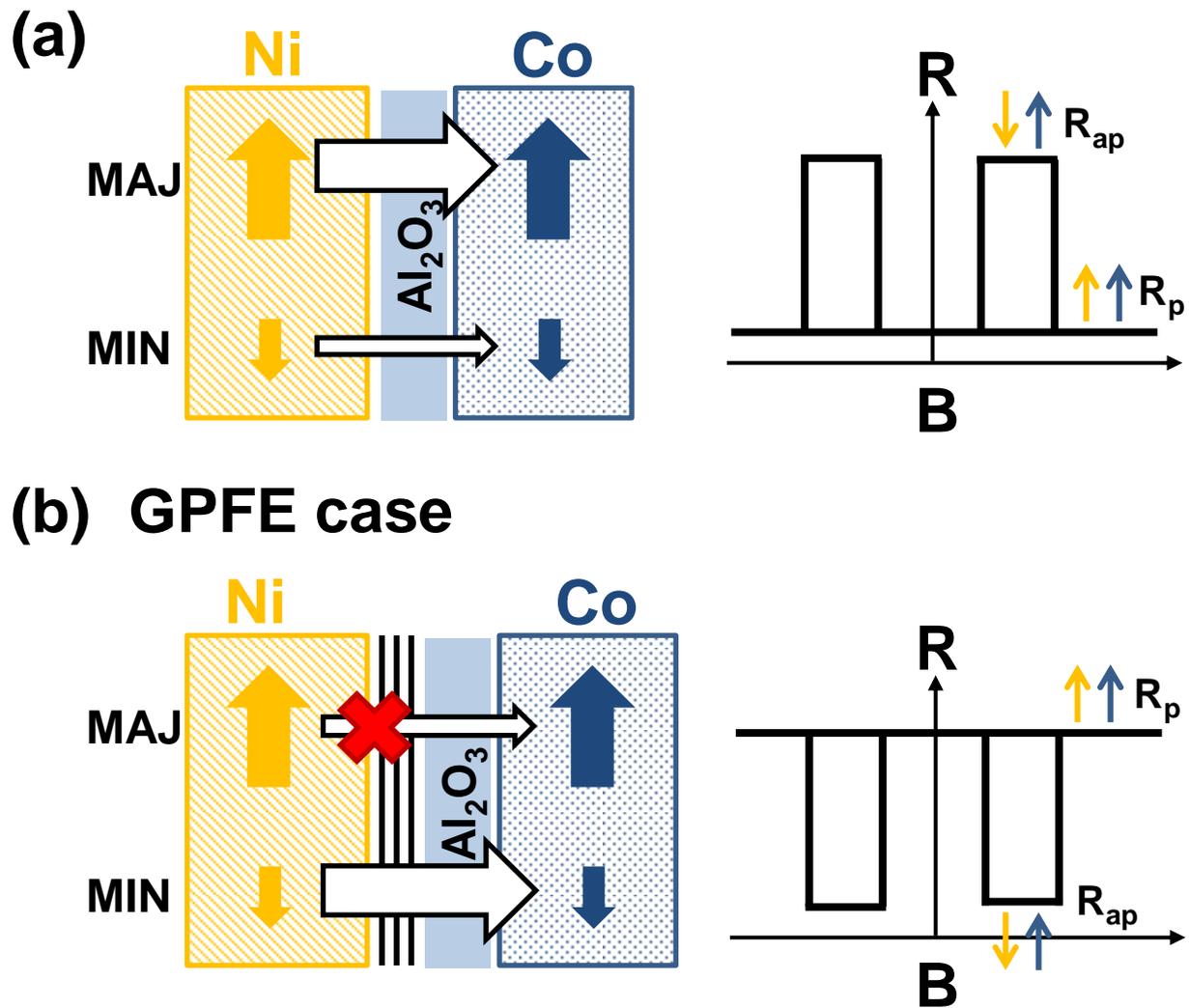
Figure 5